**Title:** Pulse-Width-Specific Phase Space Informed Universal Beam Modeling for Mobetron UHDR in FLASH-RT

Rafael Carballeira[1], David J. Gladstone [1,2], Kevin J. Willy[1], Philip Von-Voigts Rhetz[4], Rongxiao Zhang[1,3]

Thayer School of Engineering, Dartmouth College, Hanover, New Hampshire[1]

Dartmouth Cancer Center, Lebanon, New Hampshire[2]

School of Medicine, University of Missouri, Columbia, Missouri[3]

IntraOp Medical Corporation, Sunnyvale, California[4]

**Abstract**

**Objective:** FLASH radiotherapy requires ultra-high dose rates (>40 Gy/s), yet commercial treatment planning systems are currently unavailable. While Monte Carlo (MC) simulation is the gold standard for accuracy, full-head modeling is computationally prohibitive for clinical workflows. This work establishes a methodology for generating pulse-width-specific phase space files for the Mobetron UHDR (9 MeV), accounting for systematic beam quality shifts caused by RF waveguide loading across pulse widths of 1.2–4.0 µs.

**Approach:** Using GAMOS 6.2.0 MC simulations, voxelized dose distributions were post-processed via an iterative script to refine source-level parameters against experimental targets. Mean energy was optimized by matching phantom-measured R50 in the fall-off region, while energy spread was refined using surface dose and build-up gradients to capture low-energy spectral content. Relationships derived from a mid-range 6 cm diameter applicator were applied across all clinical apertures (2.5–10 cm) to test the hypothesis that beam loading occurs upstream of downstream collimation. The geometric mean of experimental pulse widths (2.28 µs) was evaluated as a universal clinical reference.

**Main Results:** Phase space files were generated for all pulse widths and apertures. For the A6I6 reference, mean energy decreased exponentially from 9.58 to 9.04 MeV ($R^2$=0.99) as pulse width increased, while energy spread increased quadratically ($R^2$=0.99), showing a strong negative correlation (r = -0.98). Cross-aperture validation confirmed the aperture-independence of these energy shifts. The universal 2.28 microsecond reference yielded a 9.32 MeV mean energy. Across experimental extremes, R50 deviations were ≤ 1.3 mm and critical depth-dose parameters (e.g., R90, R10) remained ≤ 2.0 mm, meeting AAPM TG-106 tolerances.

**Significance:** This work provides the first pulse-width-specific characterization for Mobetron FLASH planning. Validated regression models enable beam parameter prediction at arbitrary pulse widths, and the universal 2.28 μs reference reduces the computational burden by 75% while maintaining clinical accuracy.

**1. Introduction**

FLASH radiotherapy represents a paradigm shift in radiation oncology, delivering dose at ultra-high dose rates exceeding 40 Gy/s that have demonstrated significant normal tissue sparing while maintaining equivalent tumor control compared to conventional radiotherapy in preclinical studies (Favaudon et al 2014, Montay-Gruel et al 2017). This suggests a potential therapeutic window, though the clinical translation of FLASH and the scope of its advantage remain active areas of investigation. Many FLASH preclinical and early clinical investigations have used 9 MeV electron beams, making this energy regime particularly important for methodology development and treatment planning system validation.

A critical barrier to clinical translation remains: validated treatment planning systems (TPS) for FLASH delivery do not exist commercially. This is not merely a software limitation, but rather a fundamental radiation transport problem requiring accurate source characterization for dose calculation. Monte Carlo (MC) simulation is the gold standard for radiation transport in radiotherapy (Ma and Jiang 1999), but full MC simulation of the linear accelerator (linac) head and downstream delivery components for every treatment scenario is computationally prohibitive for clinical workflows. Phase space files recorded at a downstream reference plane address this by decoupling upstream beam physics from patient-specific dose calculations, enabling efficient beam modeling across different treatment geometries (Chetty et al 2007). Conventional radiotherapy TPS relies on comprehensive beam data commissioning following established protocols (Das et al 2008), but FLASH-specific commissioning frameworks remain under development.

The Mobetron, a purpose-built mobile intraoperative electron linac, has demonstrated capability for FLASH dose rate delivery in its 9 MeV ultra-high dose rate (UHDR) mode across pulse widths of 1.2–4.0 μs. To accommodate varying treatment field sizes and depth-dose requirements, the Mobetron UHDR is equipped with two interchangeable collimating applicator cones that attach to the linac head exit window, each compatible with a series of circular aperture inserts (Figure 1). The short cone applicator (19 cm length) accommodates seven aperture inserts ranging from 2.5 to 6.0 cm diameter (Figure 2), while the long cone applicator (28 cm length) accommodates five aperture inserts ranging from 6.0 to 10.0 cm diameter in 1.0 cm increments (Figure 3). Applicator configurations are designated using the notation AXIY, where X indicates the cone series (6 for short cone, 10 for long cone) and Y specifies the aperture diameter in centimeters. These geometric variations influence lateral charged particle equilibrium, beam penumbra, and output factors, motivating comprehensive phase space characterization across all clinically available configurations.

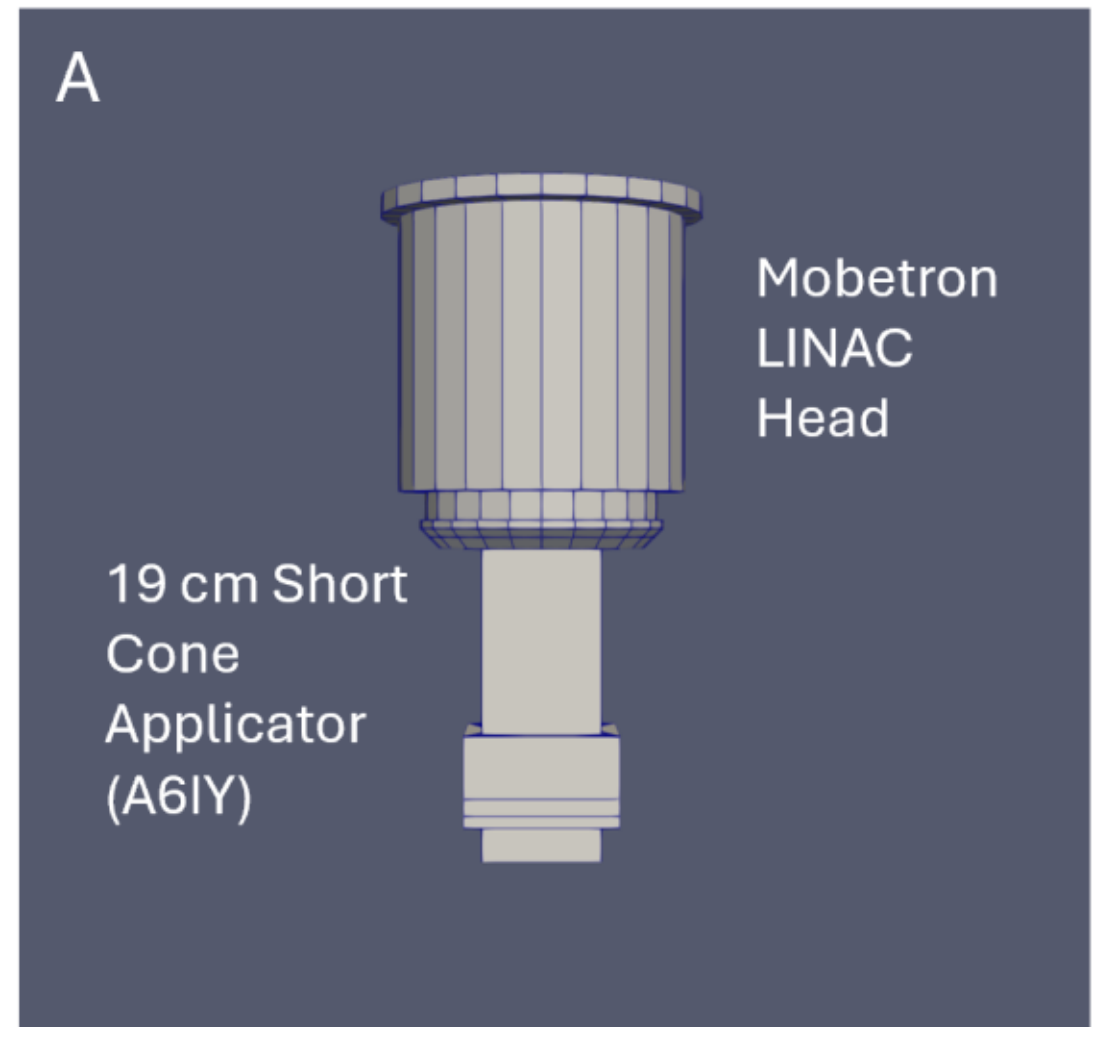

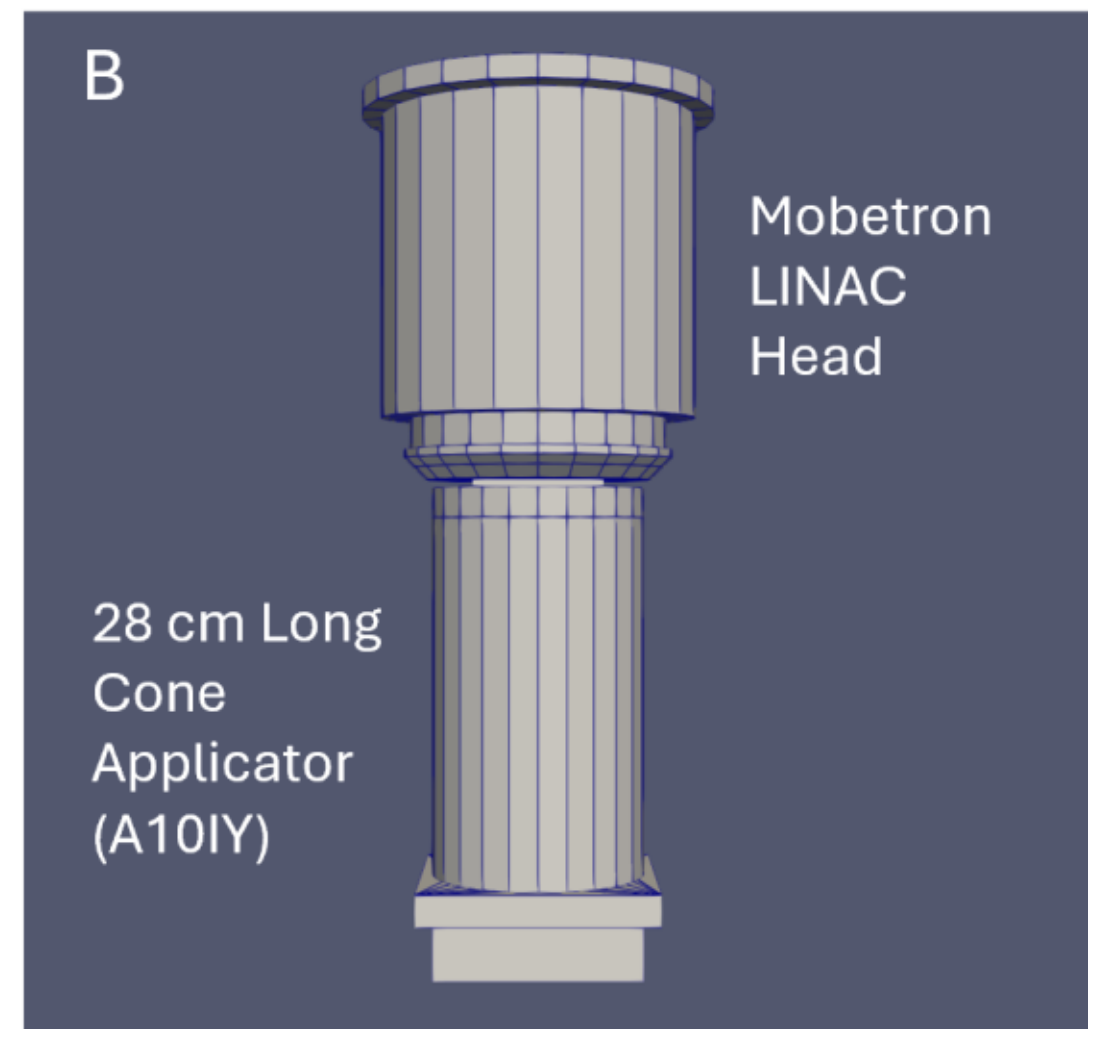

**Figure 1A-B:** Short Cone and Long Cone visualizations.

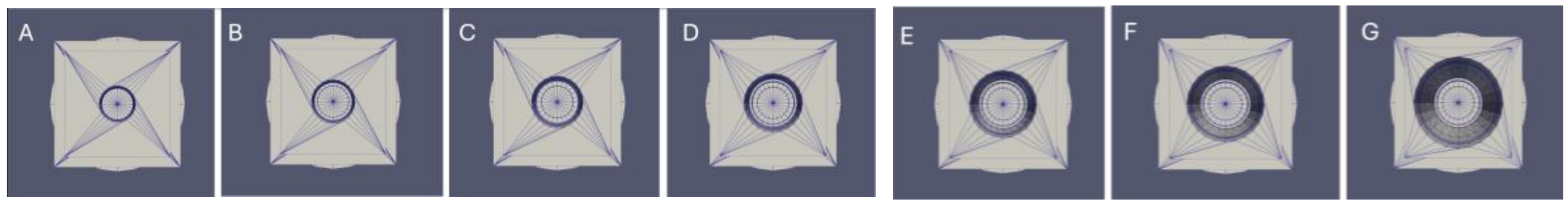

**Figure 2A-G:** A6IY configurations with aperture size, Y, following A) 2.5, B) 3, C) 3.5, D) 4, E) 4.5, F) 5, and G) 6 cm.

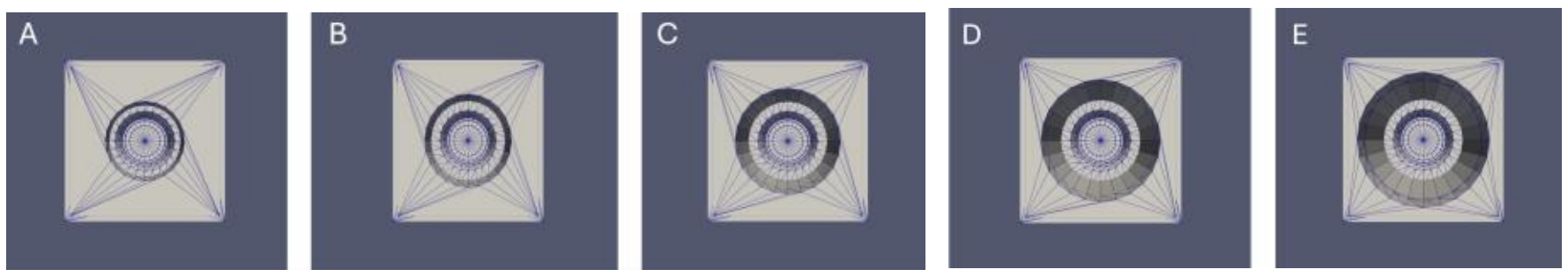

**Figure 3A-E:** A10IY configurations with aperture size, Y, following A) 6, B) 7, C) 8, D) 9, E) 10 cm.

For FLASH delivery, the standard commissioning approach is complicated by pulse width variations that introduce systematic changes in beam energy spectra through loading effects in the RF waveguide. Previous experimental characterization of the Mobetron UHDR by our group demonstrated linear relationships between pulse width and both depth of maximum dose ($d_{max}$) and half-value depth (R50), with R50 decreasing from approximately 41.6 mm at 1.2 µs to 38.0 mm at 4.0 µs—representing nearly 1 MeV of mean electron energy degradation (Dai et al 2024). Independent commissioning of the Mobetron UHDR has also been reported by Palmiero et al (2025). More recently, Audet et al (2025) characterized the Mobetron energy spectrum in greater detail and demonstrated non-Gaussian features in the lower-energy tail, most prominent in the 6 MeV mode. While the present work uses a Gaussian energy spectrum as a tractable approximation for the 9 MeV mode—where non-Gaussian effects are reported to be minimal—the iterative optimization framework presented here is spectrum-agnostic and compatible with more detailed energy distributions such as PESO-based models.

This work presents a methodology that: (1) establishes an iterative optimization process for phase space parameter refinement using experimentally validated energy–pulse width relationships, (2) validates this approach across the complete Mobetron UHDR aperture range, and (3) provides practical guidance for TPS implementation including optimal universal pulse width selection.

## 2. METHODS

### 2.1 Experimental Foundation and Initial Parameter Estimation

Starting parameters for Monte Carlo phase space generation were derived from previously published commissioning data for the Mobetron UHDR (Dai et al 2024), which characterized pulse-width-dependent beam penetration across the UHDR operating range. The Mobetron UHDR system delivers 9 MeV electron beams with customizable pulse width (0.5–4.0 μs) and pulse repetition frequency (5–120 Hz), providing flexibility in temporal beam structure while maintaining the nominal beam energy. In the commissioning study by Dai et al (2024), four base geometric configurations were characterized: 'Pristine' (no applicator, 6 cm internal shielding), 'No applicator' (inserts only, 2.5–6 cm), '6 cm applicator' (19 cm cone with 2.5–6 cm inserts), and '10 cm applicator' (28 cm cone with 6–10 cm inserts).

Briefly, that commissioning work employed EBT-XD radiochromic film calibrated against a clinical linac (0.5–40 Gy range) and positioned vertically in a custom 3D-printed PLA-walled water tank for percentage depth dose (PDD) scans (Dai et al 2024). Films were irradiated at a 1 cm depth reference point (chosen due to minimal surface build-up across most configurations) and used to acquire lateral profiles and output factors for all geometric setups. Commissioning conditions used 1.2 μs pulse width and 30 Hz PRF with a 2 cm air gap between the applicator exit and the phantom surface (4.5 cm for the pristine, uncollimated beam). Multiple measurements (n=6) were averaged to reduce film dosimetry uncertainty. Commissioning measurements were performed at 1.2 μs pulse width; this served as the reference condition for our iterative phase space optimization.

Pulse width dependence studies by Dai et al (2024) revealed that PRF had negligible effect on beam quality, while pulse width showed systematic linear relationships with penetration parameters: $d_{max}$ =

−2.08(±0.59) × PW + 23.26(±1.05) mm and R50 = −1.28(±0.09) × PW + 43.19(±0.21) mm. These empirical relationships reflect beam loading physics—as pulse width increases, greater total charge depletes RF power from the accelerating waveguide (Kim 2012, Arai et al 1980), resulting in softer energy spectra with reduced penetration depth.

To translate R50 measurements into Monte Carlo phase space parameters, we applied the established energy–range relationship for electrons in water (ICRU Report 35 1984): Mean Energy (MeV) = 2.33 × R50 (cm). A corresponding empirical relationship between pulse width and energy spread was not available in the published literature to our knowledge. We therefore derived a simple linear initialization relationship for energy spread, anchored to the Dai et al commissioning baseline of 0.20 MeV at 1.2 µs pulse width and scaled linearly at 0.05 MeV per microsecond: Energy Spread (MeV) = 0.20 + 0.05 × (PW − 1.2). This initialization served only as a starting point; the iterative optimization process described in Section 2.3 refined the energy spread independently for each pulse width based on measured surface dose and build-up gradient, such that the final optimized parameters do not depend sensitively on the initialization formulation.

**2.2 Monte Carlo Simulation Configuration**

All simulations were performed using GAMOS 6.2.0 (Geant4-based Architecture for Medicine-Oriented Simulations) (Arce et al 2014) with the standard Geant4 electromagnetic physics list. The simulation geometry was defined to begin at the source plane, located immediately downstream of the accelerating section exit, and included all passive beam delivery components: the primary scattering foils (Kapton/gold/Kapton sandwich with stainless steel housing), ion chambers, internal beam-shaping and collimating apertures, lead shielding, the interchangeable applicator cone, and the aperture insert. The electron gun, RF accelerating waveguide, and bending magnet upstream of the source plane were not explicitly modeled; their net effect on the emergent beam is represented through the source phase space parameters (mean energy, energy spread, spatial distribution), which are the quantities refined by the iterative optimization process (Section 2.3).

The source was modeled as a Gaussian energy distribution centered on the pulse-width-dependent mean energy, with a Gaussian spatial distribution (σ = 1 mm) at the source plane and a fixed direction along the beam axis. No angular divergence was applied at the source; lateral scatter is generated entirely by downstream Coulomb interactions in the passive beam-shaping elements, applicator, air gap, and phantom. This approximation is consistent with the narrow angular divergence (<0.3°) typical of clinical electron linacs (Das et al 2008, Faddegon et al 2009) and is refined implicitly through matching of measured lateral dose profiles.

Phase space files were recorded 1 mm downstream of the aperture insert exit face (z = 398 mm for the short-cone A6-series applicator, z = 479 mm for the long-cone A10-series applicator, measured from the source plane). This small offset into the air gap was adopted after testing revealed that scoring directly at the material interface produced artifacts resulting in premature depth-dose falloff during downstream transport. A water phantom ($50 \times 50 \times 20$ cm$^3$) was positioned with a 2 cm air gap between the applicator exit and the phantom surface, matching the commissioning setup of Dai et al (2024). Dose scoring used a $500 \times 500 \times 200$ voxel matrix (1 mm$^3$ resolution) throughout the phantom volume. A minimum of $10^7$ primary histories were simulated per configuration to achieve adequate statistical precision in the depth-dose and profile analyses.

**2.3 Iterative Optimization Process**

The core methodology of this work is a systematic parameter adjustment strategy that exploits the distinct physical signatures of mean energy and energy spread in different PDD regions. Because GAMOS itself does not natively support closed-loop optimization of source parameters, iteration was implemented as an external post-processing workflow: each GAMOS simulation produced a voxelized dose distribution which was then read by a custom Matlab script that computed R50, surface dose, and the build-up gradient from the simulated PDD, compared these quantities to the measured values from Dai et al (2024), adjusted the source mean energy and energy spread accordingly, generated a new GAMOS input macro, and re-ran the simulation. This loop continued until convergence criteria were met.

Mean energy adjustments were driven by R50 matching in the fall-off region, since R50 is directly proportional to mean electron energy via continuous slowing down (approximately 0.2 MeV per 1 mm R50 shift for 9 MeV electrons in water). Energy spread adjustments were driven by the build-up region, specifically the surface dose and build-up gradient, since broader energy spreads increase the low-energy electron content of the spectrum, raising surface dose and reducing gradient steepness. The script adjusted mean energy in 0.1 MeV steps when $|\Delta R50|$ exceeded the target tolerance, and refined energy spread in 0.02 MeV increments based on surface dose and gradient discrepancies. A convergence target of $|\Delta R50| \leq 0.5$ mm was initially set; in practice, convergence to within 1.5 mm was achieved for all configurations, with the largest residual deviations at the shortest pulse widths (discussed in Section 4). Decoupling the mean-energy and energy-spread adjustments through their distinct dosimetric signatures allows the iteration to converge to physically interpretable parameters rather than arbitrary curve-fit values and typically required three to five iterations per configuration.

**2.4 Profile Extraction and Validation**

Lateral dose profiles were extracted from the 3D dose matrix at the measured $d_{max}$ depth for each configuration. Profile smoothing was applied using Gaussian convolution (kernel $\sigma$ = 11 voxels for most configurations, adjusted up to 15 voxels for the largest apertures to balance noise reduction with penumbra preservation) to reduce Monte Carlo statistical noise while retaining field-edge sharpness.

**2.5 Cross-Aperture Validation Strategy**

Following successful A6I6 optimization, the derived energy-pulse width relationships were applied directly to all remaining clinical aperture configurations (2.5–10 cm diameter, spanning both short-cone and long-cone applicators) without re-optimization of the source energy parameters. This tested the hypothesis that pulse-width-dependent beam loading occurs upstream of the modeled geometry—in the RF accelerating structure itself—and is therefore independent of downstream collimation. Under this hypothesis, the same source-level energy parameters should reproduce measured depth-dose distributions across all aperture configurations, with geometric differences between apertures being captured entirely by the explicitly modeled downstream components (scattering foils, collimators, applicators, inserts).

**2.6 Statistical Analysis**

For each aperture configuration, optimized mean energy and energy spread values were analyzed as functions of pulse width using linear and exponential regression for mean energy, polynomial (quadratic) regression for energy spread, and Pearson correlation for the relationship between mean energy and energy spread. Relative energy spread was calculated as (Energy spread / Mean energy) × 100%. Given that fitting was performed over four experimental pulse width points (1.2, 2.0, 3.0, 4.0 µs), model selection prioritized physically motivated functional forms—exponential decay for mean energy (reflecting asymptotic beam loading saturation) and quadratic growth for energy spread—rather than maximal fit complexity. 95% confidence intervals and residual error metrics ($R^2$, RMSE) are reported alongside each fit.

**2.7 Universal Reference Pulse Width**

Following cross-aperture validation of the pulse-width-dependent beam parameter relationships, a single universal reference pulse width was evaluated as a practical alternative to maintaining pulse-width-specific phase space files for every clinical delivery condition. The geometric mean of the four experimental pulse widths was selected as the reference: (1.2 × 2.0 × 3.0 × 4.0)^(1/4) = 2.28 μs, compared to the arithmetic mean of 2.55 μs. The geometric mean was chosen because it corresponds to the central value on a logarithmic axis, which is the natural axis for the observed exponential mean-energy decay (Section 3.1); arithmetic averaging of pulse widths corresponds to equal weighting on a linear axis and therefore biases toward the higher-PW (lower-energy) end of the exponential curve.

The validated regression models for mean energy (E = 8.616 + 1.348 × exp(−0.284 × PW)) and energy spread ($\sigma = 0.00177 \times PW^2 + 0.0323 \times PW + 0.130$) were evaluated at 2.28 μs to predict the universal source parameters. A Monte Carlo phase space file was then generated at this reference condition for the A6I6 configuration. Validation compared the 2.28 μs universal simulation against measured depth-dose data at the experimental extremes (1.2 μs and 4.0 μs). Quantitative validation metrics included the R90, R80, R50, R20, and R10 depth parameters, together with surface dose and $d_{max}$ as build-up-region characteristics. The universal PW framework is intended as a practical simplification for routine clinical treatment planning; pulse-width-specific phase space files remain available for research applications or clinical scenarios requiring higher precision at extreme pulse widths.

## 3. Results

### 3.1 Pulse-Width-Dependent Beam Loading Validation

Iterative optimization of Monte Carlo phase space parameters reproduced the measured depth-dose distributions across all four pulse widths (1.2, 2.0, 3.0, 4.0 μs) for the A6I6 aperture configuration. Beam current transformer (BCT) measurements confirmed pulse width stability across all configurations, with coefficients of variation below 0.6%, supporting the reproducibility of the nominal pulse width settings during the commissioning experiments used as the optimization targets. F Figure 4 shows agreement between measured (solid lines) and simulated (dashed lines) PDDs and transverse profiles at d_max across all pulse width conditions. R50 deviations for A6I6 were 1.3, 0.6, 0.8, and 0.3 mm at pulse widths of 1.2, 2.0, 3.0, and 4.0 μs respectively, with the largest residual at the shortest pulse width. Profile comparisons agreed within the field boundaries (±3 cm), with field edges matching within approximately 1 mm across all pulse widths.

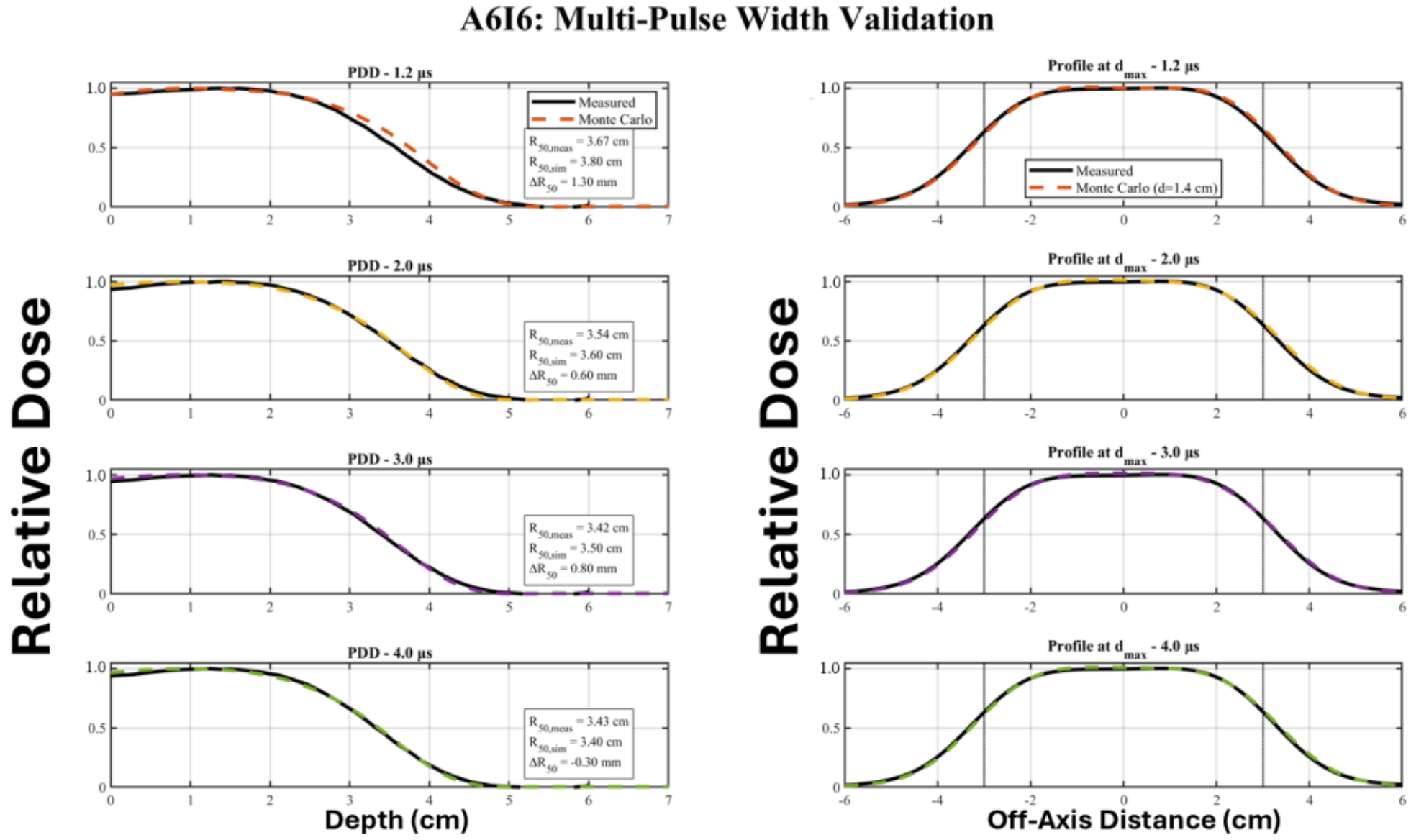

**Figure 4:** A6I6 depth-dose and profile validation across 1.2, 2.0, 3.0, and 4.0 µs pulse widths. Measured data (solid lines), Monte Carlo simulations (dashed lines).

The optimized phase space parameters revealed systematic energy variations with pulse width (Figure 5A–E). Mean energy decreased exponentially from 9.58 MeV at 1.2 µs to 9.04 MeV at 4.0 µs, following $E = 8.616 + 1.348 \times \exp(-0.284 \times PW)$ with $R^2 = 0.9946$ and RMSE = 0.0146 MeV (Figure 5A). The exponential model provided a better fit than linear ($R^2 = 0.9501$) or logarithmic ($R^2 = 0.9803$) alternatives. Energy spread increased from 0.173 MeV at 1.2 µs to 0.287 MeV at 4.0 µs, following $\sigma = 0.00177 \times PW^2 + 0.0323 \times PW + 0.130$ with $R^2 = 0.9974$ (Figure 5B). Given the sparse sampling (four data points), 95% confidence intervals for both fits are shown in the corresponding panels. The inverse relationship between mean energy and energy spread produced a strong negative correlation ($r = -0.98$, $p = 0.018$, Figure 5C), consistent with longer pulse widths simultaneously reducing mean energy and broadening the spectrum through beam loading. Relative energy spread increased from 1.81% at 1.2 µs to 3.17% at 4.0 µs (Figure 5D), reflecting progressive beam quality degradation with increasing pulse duration. Figure 5E presents a dual-axis visualization of these opposing trends across the experimental pulse width range.

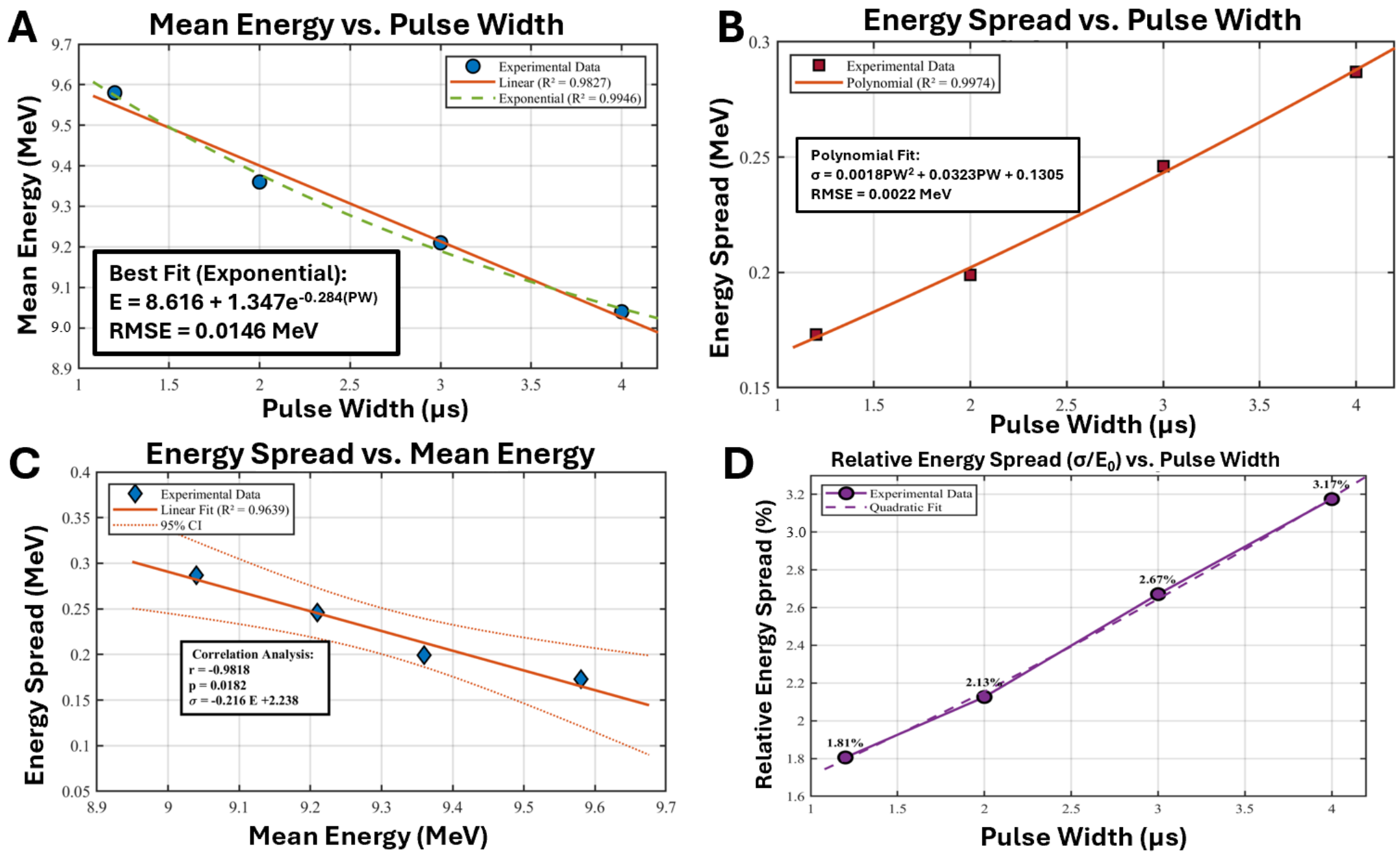

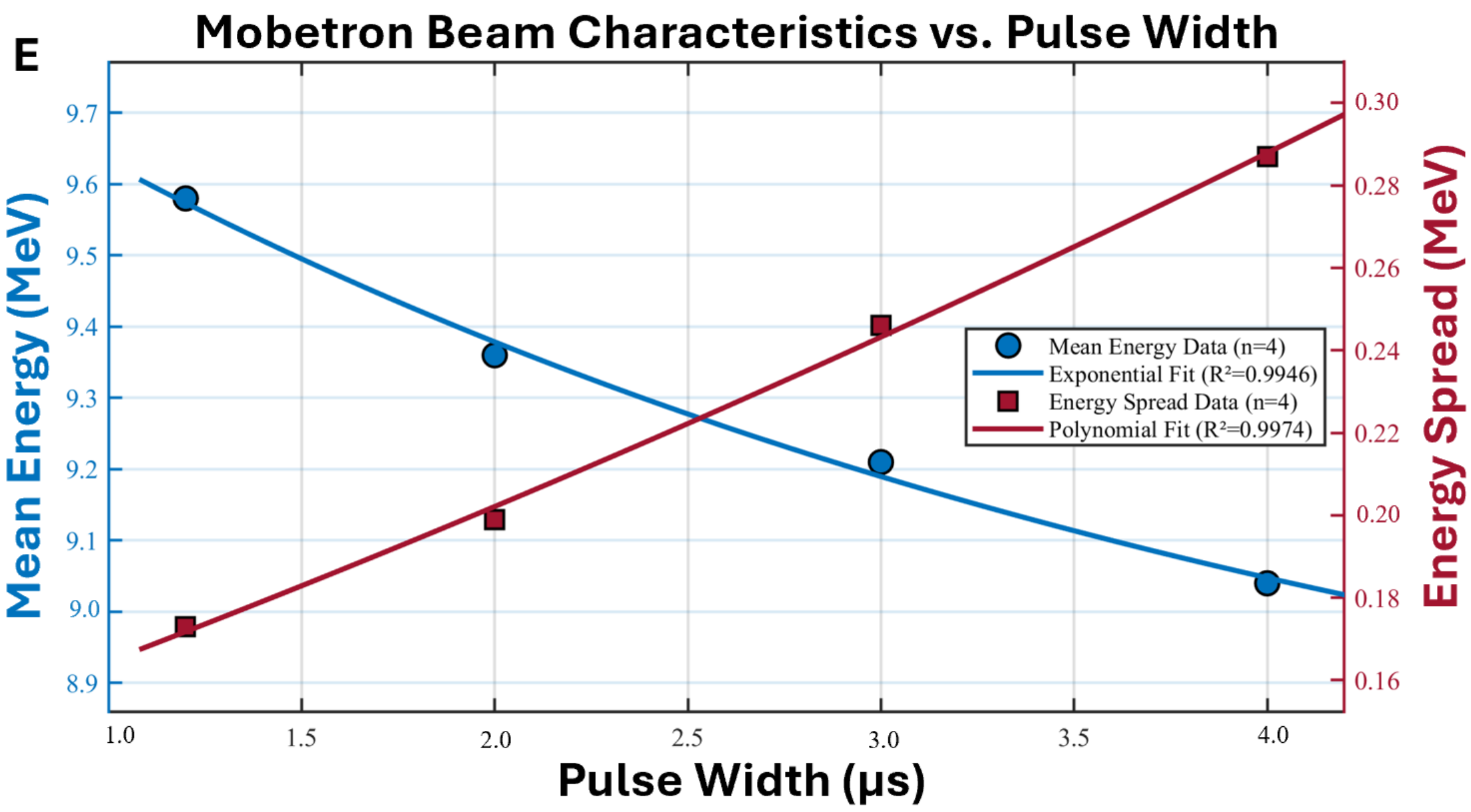

**Figure 5A-E:** Pulse-width-dependent beam parameter analysis for A6I6 configuration. A) Mean energy versus pulse width with model fits. B) Energy spread versus pulse width. C) Energy spread versus mean energy. D) Relative energy spread versus pulse width. E) Dual-axis visualization of mean energy and energy spread versus pulse width.

### 3.2 Aperture-Independent Beam Loading Physics

To test whether pulse-width-dependent beam loading depends on downstream collimation geometry, the energy–pulse width relationships derived from A6I6 were applied directly to the extreme aperture configurations (A6I25, 2.5 cm diameter; A10I10, 10 cm diameter) without re-optimization of source energy parameters. Source mean energy and energy spread were held fixed at the A6I6-derived values for each pulse width, while downstream geometry was updated to the corresponding aperture configuration. Figure 6A shows the validation for A6I25, where measured and simulated PDDs agreed with a maximum R50 deviation of approximately 1 mm across the four pulse widths. Despite A6I25 operating below the lateral scatter equilibrium radius (r ≈ 2.6 cm for 9 MeV electrons), no systematic central-axis dose deficits were observed in either the measured or simulated profiles. Figure 6B shows the validation for A10I10, the largest clinical aperture, where the maximum R50 deviation was also approximately 1 mm. Profile comparisons at $d_{max}$ showed field-edge matching within approximately 1 mm for both configurations. The applicability of A6I6-derived energy relationships across apertures from 2.5 to 10 cm diameter, without per-aperture re-optimization, supports the hypothesis that pulse-width-dependent beam loading originates upstream of the modeled geometry and is independent of downstream collimation.

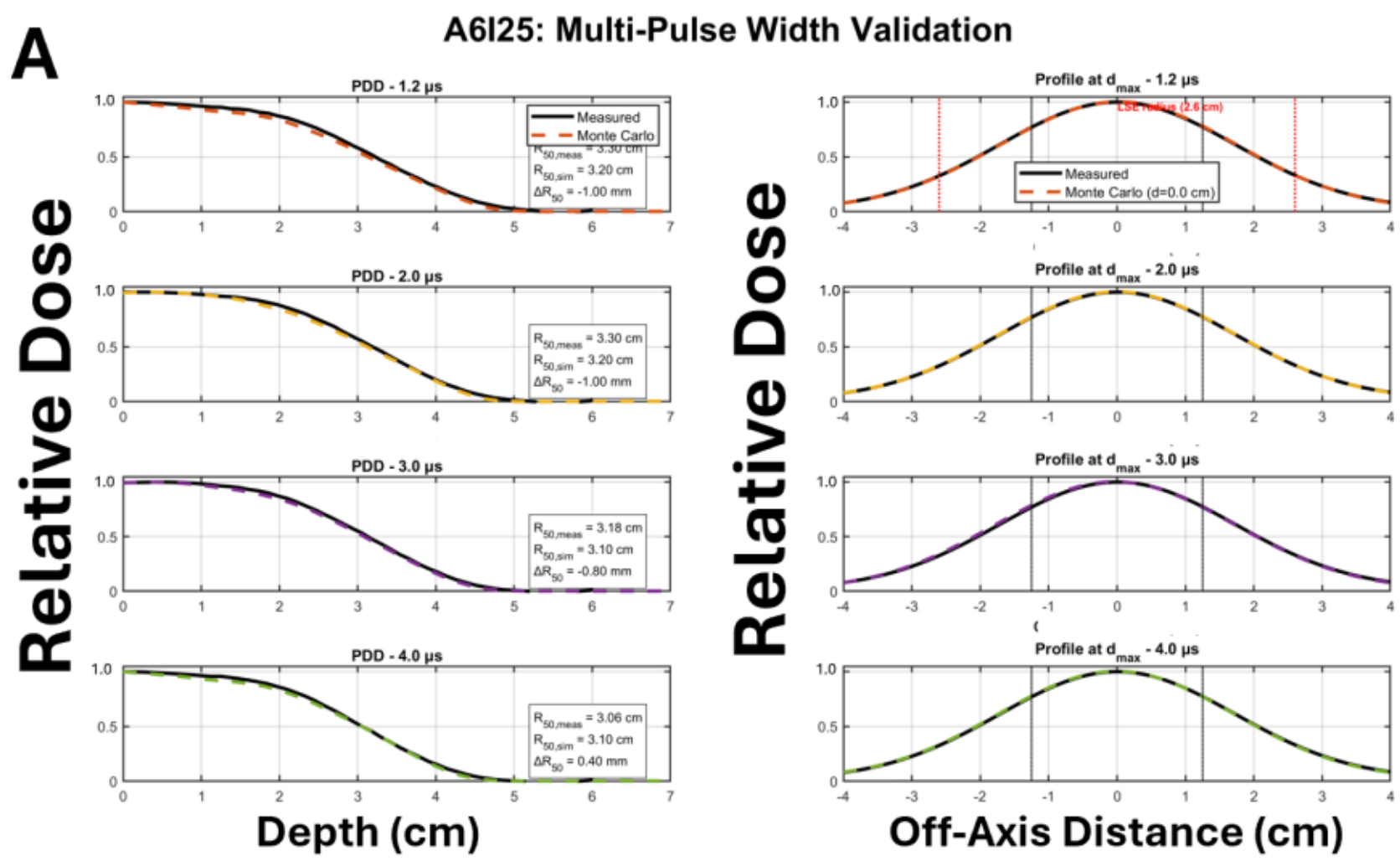

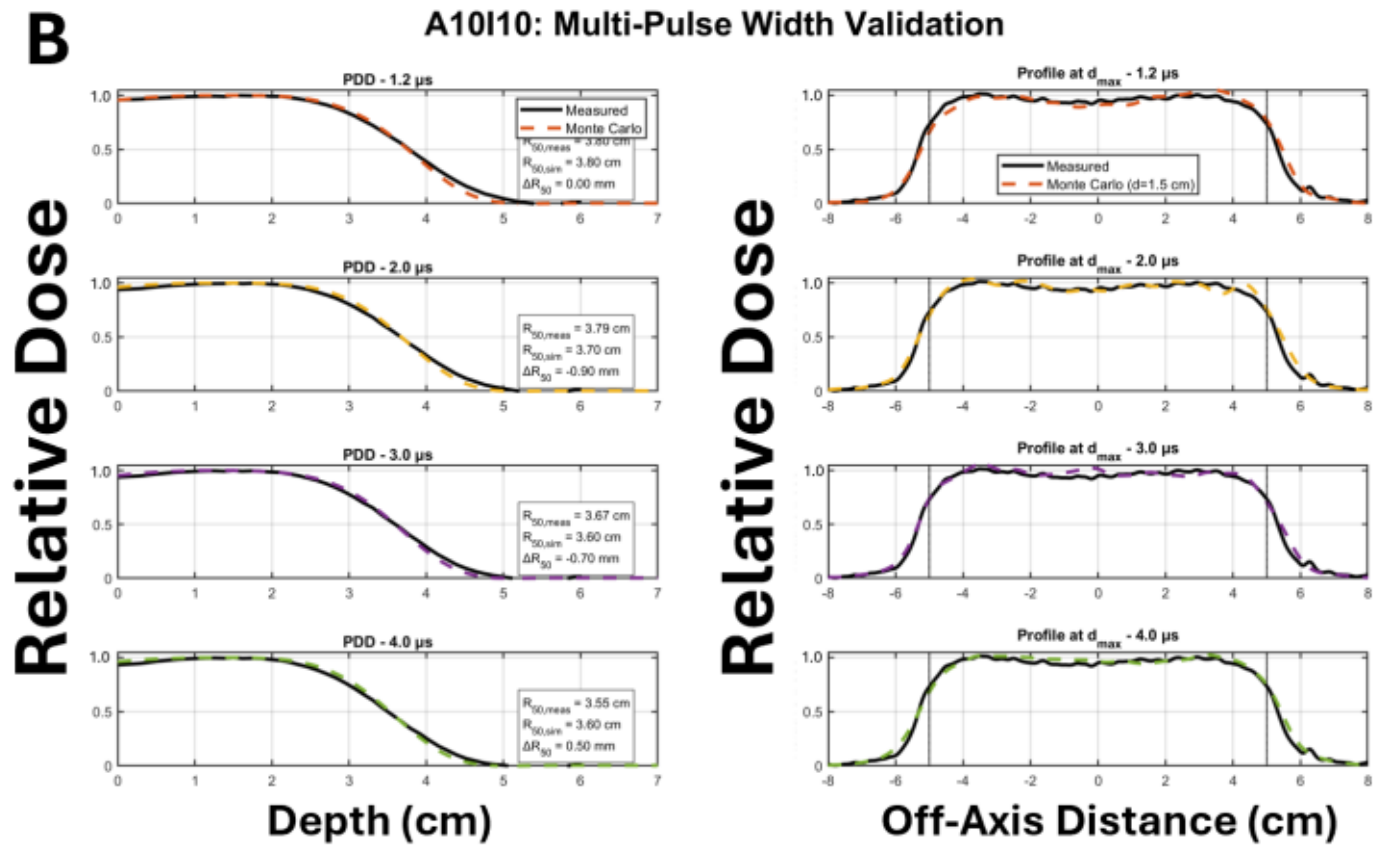

**Figure 6A-B:** Aperture-independent validation across all pulse widths. Measured data (solid black lines), Monte Carlo simulations (dashed colored lines). A) A6I25 configuration (2.5 cm diameter). B) A10I10 configuration (10 cm diameter).

### 3.3 Universal Pulse Width Validation

The geometric mean of the experimental pulse widths (2.28 µs) was evaluated as a universal reference condition. Using the validated regression models—exponential decay for mean energy ($E = 8.616 + 1.348 \times \exp(-0.284 \times PW)$, $R^2 = 0.9946$) and quadratic growth for energy spread ($\sigma = 0.00177 \times PW^2 + 0.0323 \times PW + 0.130$, $R^2 = 0.9974$)—the predicted universal source parameters at 2.28 µs were 9.32 MeV mean energy and 0.214 MeV energy spread (2.30% relative spread). This places the universal reference near the 2.0–3.0 µs operating range commonly used in FLASH protocols. Figure 7 compares the 2.28 µs universal Monte Carlo simulation against measured data at the experimental extremes (1.2 µs and 4.0 µs) for the A6I6 configuration.

Maximum absolute deviations for the critical depth-dose parameters were: R90 = 1.2 mm, R80 = 1.5 mm, R50 = 1.0 mm, R20 = 1.8 mm, and R10 = 2.0 mm, with a mean deviation of 0.8 mm across all five parameters. The 2.0 mm maximum deviation (occurring at R10) represents approximately 6% of the therapeutic range (R90–R10 ≈ 35 mm) and meets the ±2 mm tolerance for R50 recommended by AAPM TG-106 (Das et al 2008). The dose fall-off region (R80–R20), which is most relevant for target coverage and critical-structure sparing in treatment planning, showed maximum deviation below 1.5 mm.

Universal Reference Pulse Width

**Figure 7:** Universal pulse width (2.28 µs) PDD and Profile validation against experimental nominal pulse widths for A6I6 configuration. Monte Carlo simulations shown as dashed lines, measured data as solid lines.

**4. Discussion**

This work characterized pulse-width-dependent beam loading in the Mobetron UHDR and established a regression-based framework for predicting phase space parameters as a function of pulse width. The exponential decrease in mean energy and quadratic increase in energy spread observed with increasing pulse width are consistent with beam loading in the accelerating waveguide: longer pulse durations deliver greater total charge, depleting RF power and softening the resulting electron energy spectrum (Karzmark and Morton 2017, Liu F et al 2025). The validated regression models allow prediction of beam parameters at arbitrary pulse widths within the experimental range from a single commissioning exercise.

The applicability of A6I6-derived energy–pulse width relationships across all tested apertures (2.5–10 cm diameter) without per-aperture re-optimization supports the hypothesis that pulse-width-dependent beam loading originates upstream of the modeled geometry, in the RF accelerating structure itself, and is independent of downstream collimation. This has a direct practical implication: source-level phase space parameters need only be characterized as functions of pulse width, not geometric configuration, reducing the commissioning burden that would otherwise scale with the number of available applicator/aperture combinations.

While Janssen et al (2001) and Herranz et al (2015) established iterative phase space determination from measured dose data for conventional and intraoperative electron beams respectively, the present framework extends these approaches to the pulse-width-dependent beam loading regime characteristic of FLASH-capable UHDR delivery. The iterative approach exploits distinct dosimetric signatures—R50 for mean energy, surface dose and build-up gradient for energy spread—to decouple parameter adjustments into physically interpretable steps, typically converging in three to five iterations per configuration. It should be noted that the regression relationships describe the net dosimetric effect of pulse-width variation, which may include contributions beyond mean energy shift alone; Liu et al (2023) reported that intra-pulse current decreases during the pulse, introducing additional pulse-width-dependent variation in delivered dose. Because the iteration targets measured PDD features that incorporate these effects, any such contributions are implicitly captured by the fitted parameters, though they are not separated into distinct physical mechanisms by this methodology.

The geometric mean approach for selecting the universal reference pulse width (2.28 µs) is physically motivated: the geometric mean corresponds to the central value on a logarithmic axis, which is the natural axis for the observed exponential mean-energy decay, whereas arithmetic averaging biases toward the flatter high-PW region of the curve. While iterative phase space refinement is already available in conventional TPS workflows, generating four pulse-width-specific phase space files for each Mobetron UHDR installation imposes a fourfold commissioning burden and requires the selection of the appropriate phase space for each delivery condition. The regression framework presented here instead enables prediction of beam parameters at arbitrary pulse widths between 1.2 and 4.0 µs from a single commissioning exercise, and the universal 2.28 µs reference further reduces routine planning to a single phase space file while preserving the ability to generate pulse-width-specific phase spaces when precision at extreme operating conditions is required.

Several limitations of this work should be noted. First, the energy spectrum at the source plane was modeled as Gaussian. Audet et al (2025) recently characterized the Mobetron energy spectrum in greater detail and demonstrated non-Gaussian features in the lower-energy tail, most prominent in the 6 MeV mode; the present work is restricted to the 9 MeV mode where these effects are reported to be minimal. The iterative framework is spectrum-agnostic, and a more detailed spectrum model such as a PESO-based distribution could be substituted as a direct upgrade path without modifying the optimization logic. Second, regression models were fit to four experimental pulse widths; while $R^2$ values above 0.99 were obtained for physically motivated functional forms, confidence intervals (Section 3.1) reflect the sparse sampling, and extrapolation beyond the 1.2–4.0 µs range is not supported by the present data. Third, the universal 2.28 µs phase space was generated and validated for the A6I6 configuration only; the cross-aperture generalization shown in Section 3.2 uses pulse-width-specific parameters rather than the universal reference. Fourth, the "Pristine" and "No applicator" configurations from Dai et al (2024) were outside the scope of phase space file generation in this work. Finally, the methodology was developed on a single Mobetron UHDR installation; application to other installations requires verification that the factory RF tuning produces comparable beam loading behavior, since institutional tuning differences could alter the energy–pulse width relationships reported here.

The iterative optimization methodology and regression-based prediction framework may be applicable to other FLASH-capable electron linacs where pulse structure variations produce systematic energy shifts through beam loading, provided appropriate commissioning measurements are available. Validation on additional installations, extension to the 6 MeV Mobetron mode with a non-Gaussian spectrum model, and integration into a FLASH-specific TPS are natural next steps for this work.

**5. Conclusion**

This work presents a pulse-width-specific phase space characterization for the Mobetron UHDR in 9 MeV mode, developed as a step toward Monte Carlo-based treatment planning for FLASH electron radiotherapy. An iterative optimization methodology—coupling GAMOS simulations with a post-processing script that adjusts source parameters based on measured PDD features—successfully generated phase space files across the Mobetron UHDR operating range (1.2–4.0 µs), with R50 deviations within 1.3 mm across all pulse widths. Regression models captured the exponential decrease in mean energy and quadratic increase in energy spread with pulse width, enabling parameter prediction at arbitrary pulse widths within the experimental range. Cross-aperture validation at the extremes (2.5 and 10 cm diameter) supported the hypothesis that pulse-width-dependent beam loading originates upstream of the modeled geometry and is independent of downstream collimation. A universal 2.28 µs reference pulse width was validated for the A6I6 configuration, reducing routine clinical treatment planning to a single phase space file while preserving the ability to use pulse-width-specific phase spaces when precision at extreme operating conditions is required. Planned extensions include validation on additional Mobetron installations, extension to the 6 MeV mode using a non-Gaussian (PESO-based) spectrum model, and integration into a FLASH-specific treatment planning system.

**References**

1. Antolak J A, Bieda M R and Hogstrom K R 2002 Using Monte Carlo methods to commission electron beams: a feasibility study *Med. Phys.* **29** 771–86
2. Arce P, Lagares I J, Harkness L, Pérez-Astudillo D, Cañadas M, Rato P, de Prado M, Abreu Y, de Lorenzo G, Kolstein M and Díaz A 2014 GAMOS: a framework to do Geant4 simulations in different physics fields with an user-friendly interface *Nucl. Instrum. Methods Phys. Res. A* **735** 304–13

3. Audet S, Beaulieu W, Zerouali K, Guillet D, Bouchard H and Lalonde A 2025 Physics-based energy spectrum optimization (PESO): a new method to model the energy spectrum of a compact ultra-high dose rate electron linac for Monte Carlo dose calculation *Phys. Med. Biol.* **70** 085002
4. Capote R, Jeraj R, Ma C-M, Rogers D W O, Sánchez-Doblado F, Sempau J, Seuntjens J and Siebers J V 2006 *Phase-Space Database for External Beam Radiotherapy* IAEA-TECDOC-1540 (Vienna: International Atomic Energy Agency)
5. Chetty I J, Curran B, Cygler J E, DeMarco J J, Ezzell G, Faddegon B A, Kawrakow I, Keall P J, Liu H, Ma C-M C, Rogers D W O, Seuntjens J, Sheikh-Bagheri D and Siebers J V 2007 Report of the AAPM Task Group No. 105: issues associated with clinical implementation of Monte Carlo-based photon and electron external beam treatment planning *Med. Phys.* **34** 4818–53
6. Dai T, Sloop A M, Ashraf M R, Sunnerberg J P, Clark M A, Bruza P, Pogue B W, Jarvis L, Gladstone D J and Zhang R 2024 Commissioning an ultra-high-dose-rate electron linac with end-to-end tests *Phys. Med. Biol.* **69** 165028
7. Das I J, Cheng C-W, Watts R J, Ahnesjö A, Gibbons J, Li X A, Lowenstein J, Mitra R K, Simon W E and Zhu T C 2008 Accelerator beam data commissioning equipment and procedures: report of the TG-106 of the Therapy Physics Committee of the AAPM *Med. Phys.* **35** 4186–215
8. Faddegon B A, Sawkey D, O'Shea T, McEwen M and Ross C 2009 Treatment head disassembly to improve the accuracy of large electron field simulation *Med. Phys.* **36** 4577–91
9. Favaudon V, Caplier L, Monceau V, Pouzoulet F, Sayarath M, Fouillade C, Poupon M-F, Brito I, Hupé P, Bourhis J, Hall J, Fontaine J-J and Vozenin M-C 2014 Ultrahigh dose-rate FLASH irradiation increases the differential response between normal and tumor tissue in mice *Sci. Transl. Med.* **6** 245ra93
10. Guerra P, Udías J M, Herranz E, Santos-Miranda J A, Herraiz J L, Valdivieso M F, Rodríguez R, Calama J A, Pascau J, Calvo F A, Illana C, Ledesma-Carbayo M J and Santos A 2014 Feasibility assessment of the interactive use of a Monte Carlo algorithm in treatment planning for intraoperative electron radiation therapy *Phys. Med. Biol.* **59** 7159–79
11. Herranz E, Herraiz J L, Ibáñez P, Pérez-Liva M, Puebla R, Cal-González J, Guerra P, Rodríguez R, Illana C and Udías J M 2015 Phase space determination from measured dose data for intraoperative electron radiation therapy *Phys. Med. Biol.* **60** 375–401

12. Hogstrom K R and Almond P R 2006 Review of electron beam therapy physics *Phys. Med. Biol.* **51** R455–89
13. ICRU (International Commission on Radiation Units and Measurements) 1984 *Radiation Dosimetry: Electron Beams with Energies Between 1 and 50 MeV* ICRU Report 35 (Bethesda, MD: ICRU)
14. Janssen J J, Korevaar E W, van Battum L J, Storchi P R M and Huizenga H 2001 A model to determine the initial phase space of a clinical electron beam from measured beam data *Phys. Med. Biol.* **46** 269–86
15. Karzmark C J and Morton R J 2017 *A Primer on Theory and Operation of Linear Accelerators in Radiation Therapy* 3rd edn (Madison, WI: Medical Physics Publishing)
16. Liu F, Shi J, Zha H, Gao Q, Chen H and Qiu J 2025 Transient study of beam instability due to beam loading in standing-wave low-energy electron linear accelerators *Phys. Rev. Accel. Beams* **28** 050101
17. Liu K, Palmiero A, Chopra N, Velasquez B, Li Z, Beddar S and Schüler E 2023 Dual beam-current transformer design for monitoring and reporting of electron ultra-high dose rate (FLASH) beam parameters *J. Appl. Clin. Med. Phys.* **24** e13891
18. Ma C-M and Jiang S B 1999 Monte Carlo modelling of electron beams from medical accelerators *Phys. Med. Biol.* **44** R157–89
19. Montay-Gruel P, Petersson K, Jaccard M, Boivin G, Germond J-F, Petit B, Doenlen R, Favaudon V, Bochud F, Bailat C, Bourhis J and Vozenin M-C 2017 Irradiation in a flash: unique sparing of memory in mice after whole brain irradiation with dose rates above 100 Gy/s *Radiother. Oncol.* **124** 365–9
20. Palmiero A, Liu K, Colnot J, Chopra N, Neill D, Connell L, Velasquez B, Koong A C, Lin S H, Balter P, Tailor R, Robert C, Germond J-F, Gonçalves Jorge P, Geyer R, Beddar S, Moeckli R and Schüler E 2025 On the acceptance, commissioning, and quality assurance of electron FLASH units *Med. Phys.* **52** 1207–23
21. Rodrigues A, Sawkey D, Yin F F and Wu Q 2015 A Monte Carlo simulation framework for electron beam dose calculations using Varian phase space files for TrueBeam linacs *Med. Phys.* **42** 2389–403
22. Russo S, Bettarini S, Grilli Leonulli B, Esposito M, Alpi P, Ghirelli A, Barca R, Fondelli S, Paoletti L, Pini S and Scoccianti S 2022 Dosimetric characterization of small radiotherapy electron beams collimated by circular applicators with the new microsilicon detector *Appl. Sci.* **12** 600